\documentclass[prl,amssymb,twocolumn,tightenlines,showpacs,superscriptaddress]{revtex4-1}

\usepackage{graphicx}
\usepackage{amsmath}
\usepackage{verbatim}
\usepackage{subfigure}

\begin{document}
\[  \]
\title{Quantum Tomography of Inductively-Created Large Multiphoton States}

\author{E. Megidish}
\affiliation{Racah Institute of Physics, Hebrew University of
Jerusalem, Jerusalem 9190401, Israel}

\author{A. Halevy}
\affiliation{Racah Institute of Physics, Hebrew University of
Jerusalem, Jerusalem 9190401, Israel}

\author{Y. Pilnyak}
\affiliation{Racah Institute of Physics, Hebrew University of
Jerusalem, Jerusalem 9190401, Israel}

\author{A. Slapak}
\affiliation{School of Electrical Engineering, Tel-Aviv University, Tel-Aviv 69978, Israel}

\author{H. S. Eisenberg}
\affiliation{Racah Institute of Physics, Hebrew University of
Jerusalem, Jerusalem 9190401, Israel}


\begin{abstract}
The generation of quantum entangled states of many particles is a
central goal of quantum information science. Characterizing such
states is a complex task that demands exponentially large
resources as particles are being added. Previously, we
demonstrated a resource efficient source that can generate, in
principal, entanglement between any number of photons. This source
recursively fuse photon pairs generated by a pulsed laser into a
multiphoton entangled state. In the current work, we perform
quantum state tomography on the photon pair source and quantum
process tomography on the fusion operation. As a result, the full
quantum Greenberger-Horne-Zeilinger (GHZ) state of any number of
photons can be calculated. We explore the prospects of our scheme
and calculate nonlocality and genuine \textit{N}-photon
entanglement thresholds for states with up to twelve photons.
\end{abstract}

\maketitle

Multiparticle entanglement between many quantum bits (qubits) is
an important resource for quantum information science. It is
required for quantum computation
\cite{Raussendorf01,Walther05,Deutsch85} and in quantum
communication it enables error correction \cite{Schlingemann02}
and multiparty protocols, such as quantum secret  sharing
\cite{Hillery99} and open destination teleportation \cite{Zhao04}.
Multiparticle entangled states have been shown to refute local
realistic theories. The violation of these theories increases as
the particle number is increased \cite{Mermin90,Zukowski97}. These
highly entangled states can also demonstrate nonlocal interference
with a better resolution than that of the photons' fundamental
wavelength, enhancing the optical measurement accuracy
\cite{Walther04}.

The process of parametric down-conversion (PDC) in nonlinear
dielectric crystals is known to produce high-quality pairs of
polarization entangled photons \cite{Kwiat95}.  However, no
efficient higher-order process exists that can directly create
entanglement between more than two photons. One approach to create
multiphoton entangled states is to split a high-order PDC event,
in which more than one pump photon is down-converted
\cite{Kiesel07,Halevy11}. A different approach is to fuse photon
pairs produced by PDC into a multiphoton quantum state. The
largest state produced in this strategy to date is a ten-photon
GHZ state \cite{Wang16}. Usually, the produced state is
partially characterized and entanglement is verified by
measurement of an entanglement witness operator \cite{Guhne07}.

Measuring the state of a quantum system is a task of high
complexity, both experimentally and computationally. It requires
many identical copies of the system which are projected onto
different bases spanning the system's Hilbert space. The state's
density matrix is then reconstructed from the different projection
measurement results, a procedure known as quantum state tomography
(QST) \cite{James01}. If the quantum system is composed of $n$
qubits, it is $2^{n}$-dimensional, and the number of required
projection measurements is $4^{n}$. Even after sufficient amount
of data about the state has been collected, the numerical process
that is required to calculate the density matrix from the results
scales as $16^{n}$. The result of this scalability problem is that
the largest state that has been fully characterized to date is a
W-state of eight trapped ion qubits \cite{Haffner05}.

\begin{figure}
\centering
\includegraphics[angle=0,width=70mm]{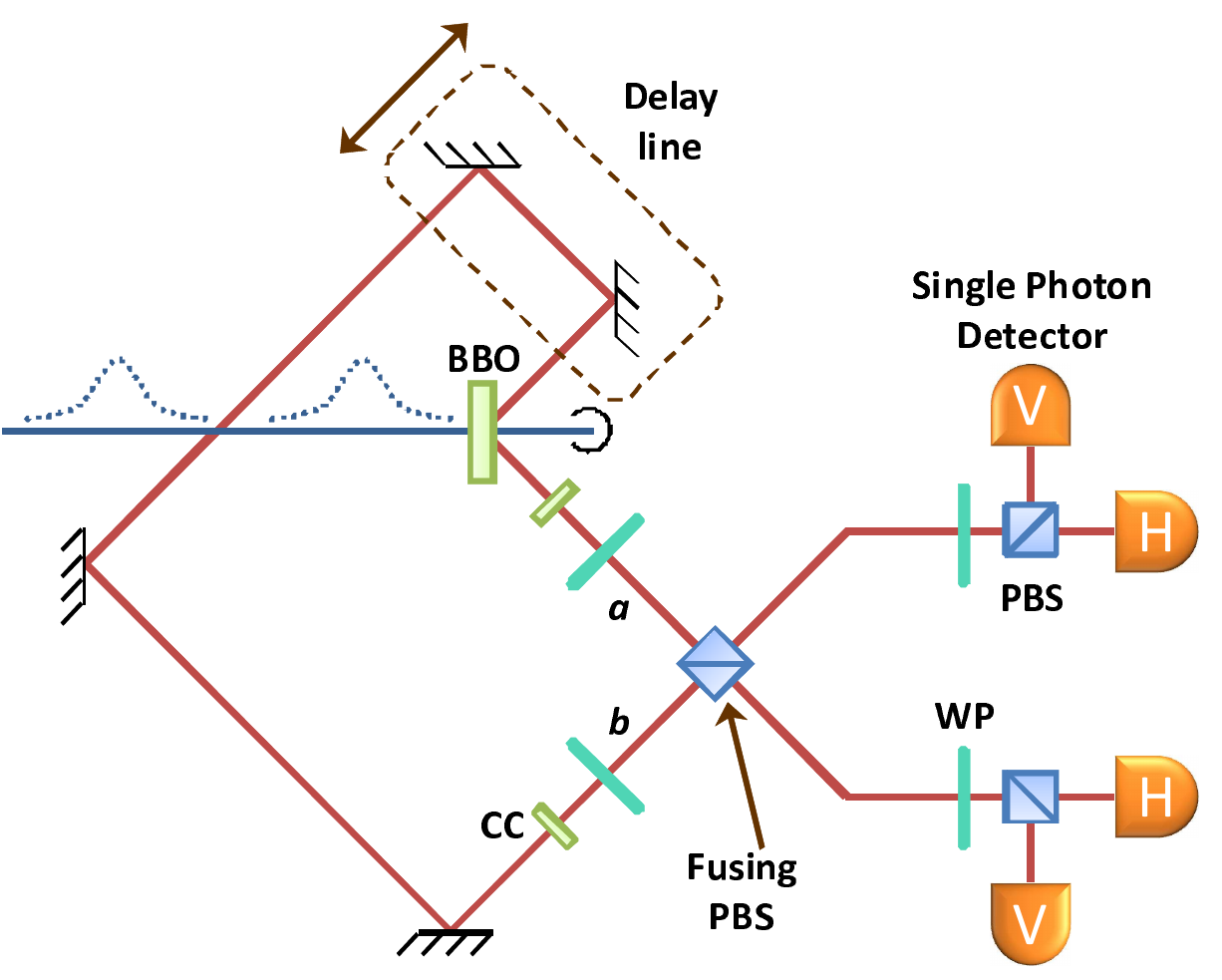}
\caption{\label{Fig1}(Color online) Experimental setup (see text for details).}
\end{figure}

We have recently introduced a resource efficient setup that can
create, in principle, entangled photon states of any number of
photons \cite{Megidish12}. A pump pulse is down-converted in a
nonlinear crystal generating pairs of  polarization entangled
photons \cite{Kwiat95}. When a pair is generated, the photon in
path \textit{a} is directed to a polarization beam splitter (PBS,
see Fig. \ref{Fig1}), while the photon in path \textit{b} enters a
delay line. The delay time $\tau$ is chosen such that if a second
entangled pair is generated by the next pump pulse, the photon of
the second pair in path \textit{a} meets the photon of the first
pair in path \textit{b} at the fusing PBS.
Post-selecting the events in which one photon exits from each PBS
output port (i.e., both photons have the same polarization),
projects the two entangled pairs onto a four-photon GHZ state as
follows
\begin{eqnarray}\label{GHZ4}
& |\psi^+\rangle^{0,0}_{a,b}\otimes|\psi^+\rangle^{\tau,\tau}_{a,b}\xrightarrow{Delay} |\psi^+\rangle^{0,\tau}_{a,b}\otimes|\psi^+\rangle^{\tau,2\tau}_{a,b}=
\\\nonumber &\frac{1}{2}(|h_a^0v_b^{\tau}\rangle+|v_a^0h_b^{\tau}\rangle)\otimes(|h_a^{\tau}v_b^{2\tau}\rangle+|v_a^{\tau}h_b^{2\tau}\rangle)\xrightarrow{PBS}
\\\nonumber  &\frac{1}{2}(|h_a^0v_b^{\tau}v_a^{\tau}h_b^{2\tau}\rangle+|v_a^0h_b^{\tau}h_a^{\tau}v_b^{2\tau}\rangle)=|GHZ\rangle_{1,2,3,4}.
\end{eqnarray}
$h_x^{t} (v_x^{t})$ is a horizontal (vertical) polarized photon
travelling in path $x$ at time delay $t$. According to this
description, time only plays a role of an additional label to each
of the photons $1,2,3,$ and $4$. Theoretically, a third pair can
be generated from the same source and its photon in path
\textit{a} will meet the delayed photon in path \textit{b} of the
second pair at the same fusing PBS. All the six photons from the
three pairs will be projected onto a six-photon GHZ state. As long
as additional consecutive pairs are generated, larger entangled
states can be produced.

In this Letter, we show that due to the recursive nature of our
scheme we are able to efficiently characterize the full density
matrix of any measured GHZ state, and even larger states that
could only be detected in principle after a very long wait. Such
states, even if they are not actually being detected, do exist in
the output amplitude of the setup with a very small probability
amplitude. For this purpose, two building blocks of the scheme are
fully characterized: the source photon pair state and the fusion
quantum process, which are sufficient for calculating the full
density matrix of any potentially-generated state. The simplicity
of this procedure enables the measurement of the full density
matrix of states of very high number of photons, which can not be
characterized using standard QST otherwise.

The created GHZ states are fused together from a set of entangled
photon pairs by several two-photon fusion operations. In the
four-photon case the final quantum state is described by the
following density matrix
\begin{equation}\label{AAPT}
\hat{\rho}_{1,2,3,4}=E(\hat{\rho}_{1,2}\otimes \hat{\rho}_{3,4})E^\dagger,
\end{equation}
where $\hat{\rho}_{i,j}$ is the photon pair density matrix of the
$i^{th}$ and the $j^{th}$ photons, and $E$ is the operator
describing the four-photon entangling process. Only photons 2 and
3 (at time $\tau$) are "interacting" at the PBS. Therefore, $E$
can be written as
\begin{eqnarray}\label{Chi}
&E=\sigma^1_0F_{2,3}\sigma^4_0,\\\nonumber
&F_{2,3}=(|h_2h_3 \rangle\langle h_2h_3|+|v_2v_3 \rangle\langle v_2v_3|)=\frac{1}{2}(\sigma^2_0  \sigma^3_0 +\sigma^2_3  \sigma^3_3),
\end{eqnarray}
where $\sigma^i_0$ and $\sigma^i_3$ are the identity and Pauli $z$
matrices when applied to the $i^{th}$ photon. In our scheme all
the pairs originate from the same source, and consequently they
are described by the same density matrix. In addition, the fusion
operation is also identical, as photons from consecutive pairs
meet at the same PBS. Thus, by measuring the two matrices
$\hat{\rho}_{1,2}$ and $F_{2,3}$, the density matrix of any
potentially-generated GHZ state can be calculated by combining
identical two-photon states with identical projections
\begin{eqnarray}
\hat{\rho}_{1,2,..,n}=&\sigma^1_0F_{2,3} \cdot\cdot\cdot  F_{n-2,n-1}\sigma^n_0\\ \nonumber
                    &(\hat{\rho}_{1,2}\otimes\cdot\cdot\cdot\otimes\hat{\rho}_{n-1,n})\\ \nonumber
                     &(\sigma^1_0F_{2,3} \cdot\cdot\cdot  F_{n-2,n-1}\sigma^n_0)^\dagger.
\end{eqnarray}
The entire information about a GHZ state containing any number of
photons is achievable, without accumulating their full statistics
or even observing them.

Polarization entangled photon pairs are created by the
non-collinear type-II PDC process \cite{Kwiat95}. A pulsed
Ti:Sapphire laser source with a $76\,$MHz repetition rate is
frequency doubled to a wavelength of $390\,$nm with an average
power of $400\,$mW. The laser beam is corrected for astigmatism
and focused on a $2\,$mm thick $\beta$-BaB$_2$O$_4$ (BBO) crystal
(see Fig. 1).  Compensating crystals (CC) correct for temporal
walk-offs. In addition, tilting of the compensating crystal in
path $a$ is used to control the phase $\varphi$ of the state
\begin{equation}
|\psi(\varphi)\rangle^{0,0}_{a,b}=\frac{1}{\sqrt{2}}(|h_a^0v_b^0\rangle+e^{i\varphi}|v_a^0h_b^0\rangle),
\end{equation}
e.g., for $\varphi= 0$ the resulting state is the maximally
entangled Bell state $|\psi^{+}\rangle$. Half-wave plates (HWP)
and quarter-wave plates (QWP) are used to analyze the photons in a
rotated basis. The $780$\,nm wavelength down-converted photons are
spatially filtered by coupling them into and out of single-mode
fibers, and spectrally filtered by using $3\,$nm wide bandpass
filters. The pair generation rate is $40,000$ per second, and the
four-photon rate is $8.5$ per second.

One photon from the first pair is delayed until another pump pulse
arrives at the generating crystal by a 31.6\,m long ($105\,$ns)
free-space delay line. The delay time enables the fusion of pairs
which are separated by eight consecutive laser pulse. This delay
time is also longer than the dead time of the single photon
detectors ($50\,$ns, Perkin Elmer SPCM-AQ4C). The delay line is
built from highly reflective dielectric mirrors, with an overall
transmittance higher than $90\%$ after $10$ reflections. Less than
$10\%$ of the signal is sampled into a single mode fiber as a
feedback signal that is used to stabilize the delayed beam's
spatial properties, by tilting a piezo-mounted mirror in the
middle of the delay line. Before any scan the delay time is
measured and calibrated using two-photon interference
measurements.

\begin{figure}
\centering\includegraphics[angle=0,width=86mm]{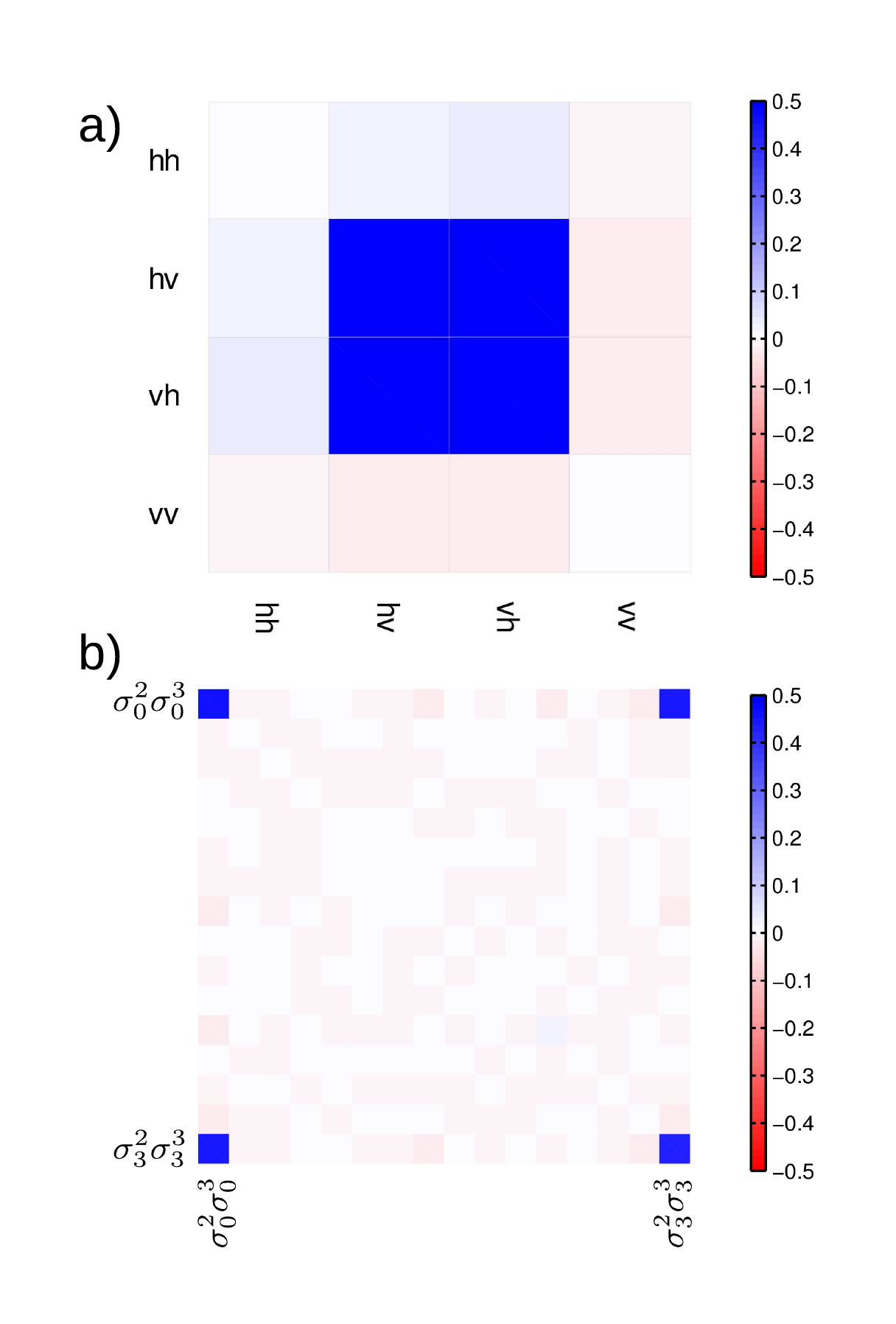}
\caption{\label{Fig2} a) Real part of the
photon pair density matrix, $\hat{\rho}_{1,2}$. The polarization
entangled pairs are generated with a $97.7\%$ overlap with the
$|\psi^+\rangle$ Bell state. b) Real part of the $\chi$ matrix
representing the fusion operation measured by AAPT technique.}
\end{figure}

First, we performed standard QST and reconstructed the pair
density matrix $\hat{\rho}_{1,2}$ \cite{James01}. When we consider
detection of only pairs, the fusing PBS serves as the polarization
analyzer (see Fig. 1). The polarization basis is controlled using
wave plates in each path ($a$ and $b$) prior to the fusing PBS. We
discriminate between the photons by their arrival times to the
detectors ($0$ and $\tau$). As a result, each photon polarization
amplitude is labelled in time and space ($0$ and $\tau$, a and b).
The two photon state, as can be seen in Fig. \ref{Fig2}(a),
was measured to have a $97.7\%$ overlap with the $|\psi^+\rangle$
state.

Next, we measured the fusing operation between photons 2 and 3. In
a standard quantum process tomography one should measure the
process output for different input states \cite{OBrien04}. We used
ancilla-assisted process tomography (AAPT) \cite{Mohseni08}, where
each input photon is entangled to another photon and a QST is
performed to all four photons. In AAPT, the input polarization
states of the quantum process are controlled by measuring the
ancillary entangled photons in different polarization bases. This
is also manifested by the Choi-Jamio{\l}kowski isomorphism between
completely positive maps and quantum states \cite{Jamiolkowski72}.
In our case, the fusion process of photons 2 and 3 can be
extracted from the initial pair density matrix,
$\hat{\rho}_{1,2}$, and by performing QST on the final four-photon
state.

Four-photon QST requires at least $256$ different projections to
span the entire four-photon polarization Hilbert space. For the
polarization manipulation of the two photons at time $\tau$ we
used WPs positioned after the fusing PBS. The polarization states
of photons 1 and 4 arriving at times $0$ and $2\tau$ are
controlled non-locally by the WPs and the biregringent phase
$\varphi$ before the fusing PBS (see supplemental material of Ref.
\cite{Megidish13} for more details). The 256 probabilities where
measured by the four detectors using 81 WPs configuration. Each
configuration was integrated for $30$ seconds and after each set
the delay line was calibrated. We repeated this sequence 78 times.

\begin{figure}
\centering\includegraphics[angle=0,width=86mm]{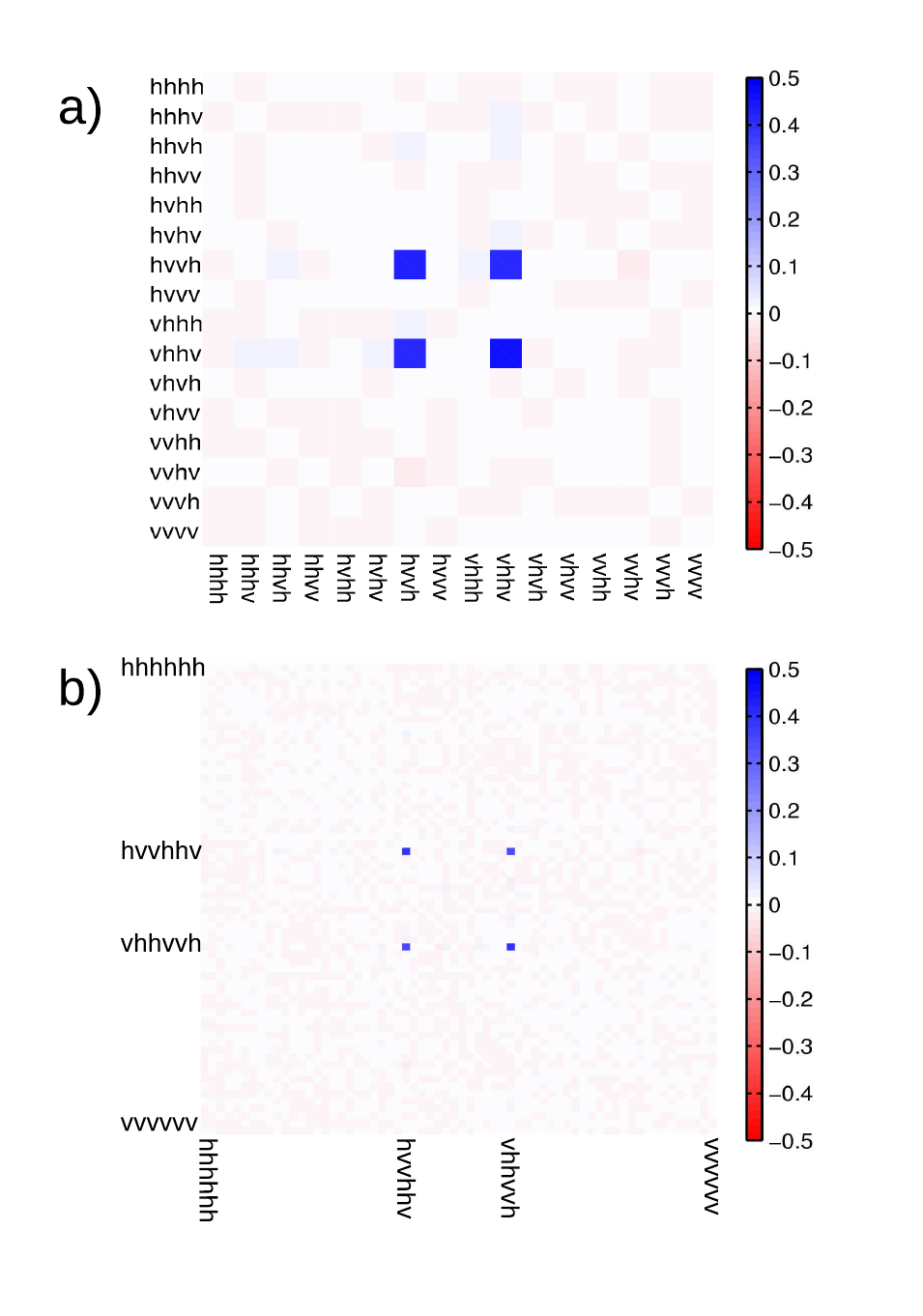}
\caption{\label{Fig3}(Color online) 
Real part of the four-photon ,
(a) and six-photon density matrices 
(b). The states were calculated from the fusion
process and the photon pair density state. 
}
\end{figure}

The process operator can be represented in the Pauli basis as
\begin{equation}\label{MaximallLiklehood}
\hat{\rho}_{1,2,3,4}=\sum_{i,j=0}^3\chi_{i,j}( \sigma^1_0\sigma^2_i\sigma^3_j\sigma^4_0 )(\hat{\rho}_{1,2}\otimes \hat{\rho}_{3,4})( \sigma^1_0\sigma^2_i\sigma^3_j\sigma^4_0 )^\dagger,
\end{equation}
where $\chi_{i,j}$ is the process coefficient matrix which
completely and uniquely describes the process.  The $\chi$ matrix
is reconstructed from the measured initial state
$\hat{\rho}_{1,2}$ and the projection measurements of the final
state $\hat{\rho}_{1,2,3,4}$, by a maximal likelihood fit to Eq
\ref{MaximallLiklehood} (see Fig. \ref{Fig2}(b)).
We then calculated the four-photon density matrix using the fusion
operation and the pair density matrix (see Fig.
\ref{Fig3}(a)). Two distinct populations ($h_1v_2v_3h_4$ and
$v_1h_2h_3v_4$) with the corresponding coherence terms are clearly observed, in accordance with the expected four-photon GHZ state of Eq .\ref{GHZ4}. The fidelity between the calculated four-photon GHZ state and the ideal one is $(85.4\pm0.2)\%$.
Furthermore, we calculated the six-photon density matrix (see Fig. \ref{Fig3}(b)). Similar to four-photon state the six-phton state also shows the GHZ characteristics with fidelity $(74.3\pm0.4)\%$.
Experimentally we detected six-photons events at a rate of 20 events per hour.

Errors were estimated with a bootstrap method, by a Monte Carlo
simulation of the fusion reconstruction process. Due to the high
flux of pairs ($40,000$ per second) the errors in the pair density
matrix were neglected. Each projection measurement of the final
state $\hat{\rho}_{1,2,3,4}$ is assumed to have a Poissonian error
distribution around the average event count. Numerically, we
created 100 randomly distributed measurement samples. For each
sample, the $\chi$ process matrix was reconstructed by the maximal
likelihood procedure. We calculated the different measures for the
entire error sample $\{\chi\}_1^{100}$ and took the standard
deviation as the error.

The four-photon fidelity is mainly affected by distinguishability
between photons 2 and 3. The photons can be distinguished
(labelled) by their arrival time, spectrum, angle, beam width, and
position. The setup was optically designed, calibrated and
actively stabilized to minimize these distinguishabilities.
Nevertheless, minute temperature changes and spectral difference
between the photons due to the PDC process \cite{Mosley} introduce
some distinguishability. In addition, the polarization of photons
1 and 4 is controlled by nonlocal rotations. Though the pair
fidelity is high ($97.7\%$), some polarization rotation error is
unavoidable and therefore some inaccuracy is introduced to the
AAPT.



\begin{figure}
\centering\includegraphics[angle=0,width=86mm]{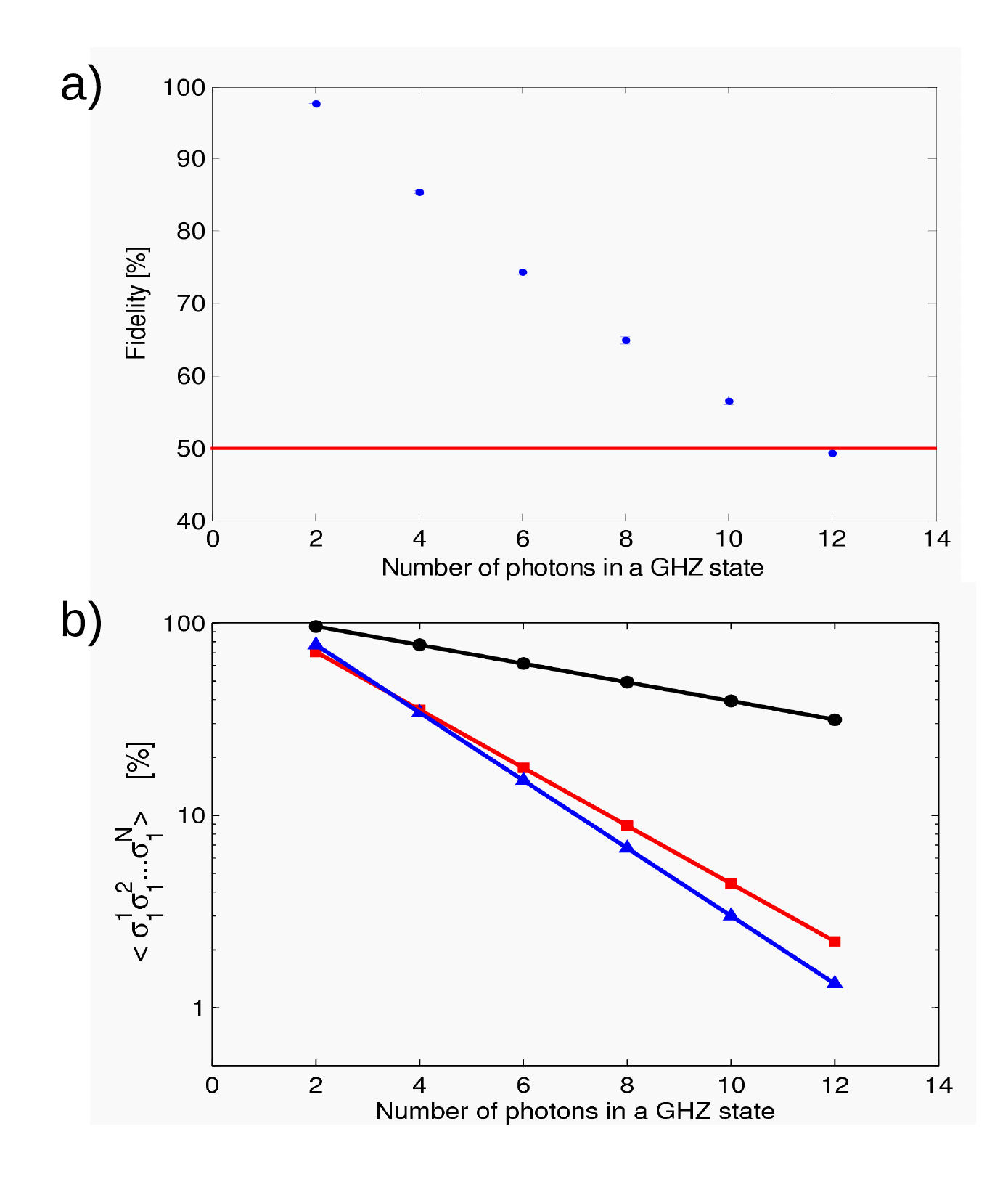}
\caption{\label{FIdVsN} (a) Fidelity between an ideal
\textit{N}-photon GHZ state and the state calculated from the pair
density matrix and the fusion operation as a function of the
number of photons. Red line indicates the $50\%$ fidelity
threshold required for genuine \textit{N}-photon entanglement.
Errors were calculated as described in the main text.
(b) \label{VisVsN} (black) Calculated interference visibility
when all photons are rotated to the $p/m$ basis as a function of
the number of photons in the GHZ state. Threshold visibility
required to violate local realism according to \.{Z}ukowski
\textit{et al.} (red) \cite{Zukowski97} and Mermin (blue)
\cite{Mermin90}.
}
\end{figure}

Using the fully characterized two-photon density matrix and the
fusion process matrix, we calculated the full density matrices of
six, eight, ten, and twelve photons, and their fidelities with the
corresponding GHZ states (see Fig. \ref{FIdVsN}). As more pairs
are added, the overlap between the \textit{N}-photon GHZ state and
the calculated state decreases due to the imperfect fusion
operation. Genuine \textit{N}-photon entanglement for GHZ states
requires fidelity above $50\%$ \cite{Sackett00}. We currently
satisfy this condition up to ten photons, where the value for
twelve photons is $(49.3\pm0.65)\%$.

Multiphoton entangled states with higher photon numbers refute
local realism with increasing violation. Bell inequalities for
\textit{N}-particle GHZ state with different number of measurement
settings set a criteria for the threshold visibility for refuting
local realistic theories \cite{Mermin90,Zukowski97}. We have
calculated the expectation value of the
$\sigma_1^1\sigma_1^2\cdot\cdot\sigma_1^n$ operator corresponding
to the visibility when the photons are measured in the $p/m=h \pm
v$ basis, and compared them to the threshold values obtained by
Mermin \cite{Mermin90} and  \.{Z}ukowski \textit{et al.}
\cite{Zukowski97} (see Fig. \ref{VisVsN}). The computed visibility
for any possibly generated GHZ state from our setup refute local
realism according to the criteria above (see Fig. \ref{VisVsN}).

In conclusion, we have measured the pairs' density matrix and the
fusion operation of our multiphoton entanglement setup. The full
quantum state of any potentially-generated GHZ state can be
calculated without even detecting the relevant photons, avoiding
the experimental and computational complexity that is usually
required for the characterization of such states. For the first
time, we can experimentally explore the different prospects of
schemes that generate multiphoton entanglement with high photon
numbers. Currently, our system has the potential of refuting local
realistic theories in any number of photons and to generate
genuine multiphoton entangled states of up to ten photons.

The authors thank the Israeli Science Foundation for supporting
this work under grants 546/10 and 793/13.

\end{document}